\documentclass[useAMS,usenatbib]{mn2e}
\usepackage{natbib}
\usepackage{graphicx}
\usepackage{color}
\usepackage{enumerate}
\usepackage{verbatim}  
\usepackage{booktabs}
\usepackage{amsmath}
\usepackage[table]{xcolor}
\usepackage[section]{placeins}
\usepackage{float}
\usepackage{enumerate}
\begin{document}
\title[Magnetic inclination angle and the red timing-noise of pulsars]{The evolution of the magnetic inclination angle as an explanation of the long term red timing-noise of pulsars}
\author[Yi \& Zhang]{Shu-Xu Yi$^{1,2}$,
Shuang-Nan Zhang$^{1,3}\thanks{E-mail:
zhangsn@ihep.ac.cn}$\\
$^{1}$ Key Laboratory of Particle Astrophysics, Institute of High Energy Physics, Chinese Academy of Sciences, Beijing 100049, China\\
$^{2}$ University of Chinese Academy of Sciences, Beijing 100049, China.\\
$^{3}$ Space Science Division, National Astronomical Observatories of
China, Chinese Academy of Sciences, Beijing 100012, China\\}

\maketitle
\begin{abstract}
We study the possibility that the long term red timing-noise in pulsars originates from the evolution of the magnetic inclination angle $\chi$. The braking torque under consideration is a combination of the dipole radiation and the current loss. We find that the evolution of $\chi$ can give rise to extra cubic and fourth-order polynomial terms in the timing residuals. These two terms are determined by the efficiency of the dipole radiation, the relative electric-current density in the pulsar tube and $\chi$. The following observation facts can be explained with this model: a) young pulsars have positive $\ddot{\nu}$; b) old pulsars can have both positive and negative $\ddot{\nu}$; c) the absolute values of $\ddot{\nu}$ are proportional to $-\dot{\nu}$; d) the absolute values of the braking indices are proportional to the characteristic ages of pulsars. If the evolution of $\chi$ is purely due to rotation kinematics, then it can not explain the pulsars with braking index less than 3, and thus the intrinsic change of the magnetic field is needed in this case. Comparing the model with observations, we conclude that the drift direction of $\chi$ might oscillate many times during the lifetime of a pulsar. The evolution of $\chi$ is not sufficient to explain the rotation behavior of the Crab pulsar, because the observed $\chi$ and $\dot{\chi}$ are inconsistent with the values indicated from the timing residuals using this model.
\end{abstract}
\section{Introduction}
Temporally correlated residuals are very common in the pulse time of arrival (TOA) of pulsars, after accounting for the standard model of spin down, astrometric variations, interstellar medium (ISM) effects and potential binary motion. The systematic deviations from the rotation model are referred as red timing-noise, since the TOA residuals have more power in the low Fourier frequencies than in the high frequencies. In order to find the signals of gravitational waves (GW) hidden in timing data \citep{Sazhin1978,Detweiler1979}, we need to model other sources of red noise as best as possible.

Most mechanisms which are proposed to explain the red timing-noise can be sorted into two classes: a) random transfer of angular momentum, either between the neutron star curst and the interior \citep{Jones1990}, or between the neutron star and the fallback matter \citep{Boynton1972} ; b) variance of the braking torque \citep{Kramer2006,Lyne2010}. A new attempt was made by \cite{Shannon2013}, who attributed the red noise in the timing residuals of PSR B1937+21 to an unknown belt of asteroids. A swarm of orbiting objects around the pulsar is equivalent in mathematics to introducing many sinusoid waves components into the timing residuals. In fact, the timing residuals of PSR B1937+21 in the time span of the study of \cite{Shannon2013} can be more naturally fitted to a single cubic polynomial rather than to a series of sinusoids \citep{Lyne2015}.

In the sample of \cite{Hobbs2010}, a considerable number of pulsars' timing residuals show the shape of a cubic polynomial, and some of the other pulsars exhibit the forth-order polynomial residuals or quasi-periodic structures, after fitting the pulse arrival phases to the spin frequency derivative ($\dot{\nu}$). If the braking of pulsars is mainly contributed by the dipole radiation of a constant magnetic field, which is a simplest assumption and is widely used, the timing model needs only to include up to the $\dot{\nu}$ term. However, the unexpectedly significant cubic and higher order terms in the timing residuals compel us to revisit such assumption.

Assume that the braking of a pulsar can be expressed as follows:
\begin{equation}
I\nu\dot{\nu}=-\frac{2\pi^2B_*^2\nu^4R^6}{3c^3},\label{first}
\end{equation}
where $I$ and $R$ are the momentum of inertia and the radius of the pulsar, and $B_*$ is the equivalent magnetic field.
We do not presume the $B_*$ to be constant, and as long as the relative change of the equivalent magnetic field $\dot{B_*}/B_*$ is small, the following expression is valid:
\begin{equation}
B_*=B_{*0}\left(1+b(t)\right),\label{four}
\end{equation}
where $b(t)$ is the dimensionless time variant part of $B_*$.

Integration of Equation (\ref{first}) twice gives the pulse arrival phase:
\begin{equation}
\Phi(t)=\Phi_0+\nu_0t-AB_{*0}^2\nu_0^3\frac{t^2}{2}+\frac{3A^2B_{*0}^4}{\nu_0}\frac{t^3}{3!}-2AB_{*0}^2\nu_0^3\iint b(t)dt^2,\label{five}
\end{equation}
where $\nu_0$ is $\nu$ at the beginning of the observation ($t=0$), $\Phi_0$ is a constant phase offset and $A\equiv 2R^6/(3c^3I)$.

From Equation (\ref{five}) one can clearly see that the linear term in $b(t)$ will give rise to an extra cubic term in the timing residuals, and the $n$-th order polynomial term in $b(t)$ will produce the $n+2$-th order polynomial term in the timing residuals. In other words:
\begin{equation}
b(t)=-\frac{\nu_0}{2\dot{\nu_0}}\frac{d^2R(t)}{dt^2},\label{useful}
\end{equation}
where $R(t)$ is the timing residuals defined as:
\begin{equation}
R(t)=\left(\Phi(t)-(\Phi_0+\nu_0t+\dot{\nu}_0\frac{t^2}{2}+3\frac{\dot{\nu}^2}{\nu_0}\frac{t^3}{3!})\right)/\nu_0,
\end{equation}
and $\dot{\nu_0}\equiv-AB_{*0}^2\nu_0^3$.

As for the evolution of $B_*$, one proposed mechanism is the decay of $B$ through Ohmic dissipation \citep{Haensel1990}. The monotonically decreasing of $B_*$ can only count for a positive $\ddot{\nu}$, whereas in observations, the number of positive-$\ddot{\nu}$-pulsars does not overwhelm that of negative $\ddot{\nu}$. In order to solve that problem, \cite{ZhangXie2012} proposed that the long term decay of $B$ is modulated by short term oscillations. Therefore, in a time span shorter than the oscillation period, the value of $B$ can be both increasing and decreasing, and the sign of $\ddot{\nu}$ can be both positive and negative. 

Another possibility was proposed by \cite{Lyne2013} that the magnetic inclination angle $\chi$ drifts, and they argued that the consequent change of the equivalent magnetic field can take account for the 45 years of timing residuals of the Crab pulsar. In this work we further study this possibility. In Section 2 and 3, we derive how the evolution of $\chi$ will affect the timing residuals, given that the braking torque is the combination of the dipole radiation and the current loss, and how this model can explain observations. In section 4, we discuss the origins of the $\chi$ evolution, and what we can learn from the observed braking index. In section 5, we discuss how our theory is related to observations, and what can not be explained by this mechanism. We conclude our work in the final section.
\section{The equivalent magnetic field when the current loss torque is present}
The magnetic dipole radiation braking mechanism predicts that $\chi$ of a pulsar approaches 0 or $180^\circ$, and if the braking is dominated by the current loss, $\chi$ approaches $90^\circ$ \citep{Barsukov2009}. As was pointed out by many authors \citep{Beskin1993,Mestel1999,Beskin2007} that the dipole braking is not efficient when the pulsar is surrounded by plasma, and the observed profile evolution of the Crab pulsar implies that $\chi$ is drifting towards $90^\circ$ as expected by the current loss mechanism \citep{Lyne2013}. Therefore, we consider the total braking torque of a pulsar as a combination of both mechanisms \citep{Jones1976}:
\begin{equation}
\mathbf{K}=\alpha\mathbf{K}_{\rm{dip}}+\mathbf{K}_{\rm{cur}},
\end{equation}
where $0<\alpha<1$ is to take account the inefficiency of dipole radiation braking. 
$\mathbf{K_{\rm{dip}}}$ is parallel with the spin axis, and its value is:
\begin{equation}
K_{\rm{dip}}=-2\pi\nu^3B^2_0A\sin^2\chi;
\end{equation}
$\mathbf{K}_{\rm{cur}}$ is in the direction of the magnetic dipole; the component parallel to the spin axis contributes to the braking of the pulsar:
\begin{equation}
K_{\rm{cur}\parallel}=-2\pi\beta\nu^3B^2_0A\cos^2\chi,
\end{equation}
where $\beta$ is a scaling factor proportional to the ratio between the electric-current density in the pulsar tube and the Goldreich-Julian current \citep[see][Equation (12)]{Barsukov2009}.

As a result of the combination, the equivalent magnetic field of Equation (\ref{first}) is
\begin{equation}
B_*^2=B_0^2\left(\alpha\sin^2\chi+\beta\cos^2\chi\right).\label{ten}
\end{equation}
Then Equation (\ref{ten}) can be rewritten as:
\begin{equation}
B_*^2=B^2_{*0}(1+\frac{\beta-\alpha}{\beta+\alpha}\cos(2\chi)),\label{ele}
\end{equation}
where $B^2_{*0}\equiv \left[(\alpha+\beta)/2\right]B^2_0$.

\section{The magnetic inclination angle evolution and the timing residuals}
It is reasonable to assume that during the observation time span, the change rate of $\chi$ can be treated as a constant $\dot{\chi}_0$, therefore
\begin{equation}
 \chi(t)=\chi_0+\dot{\chi}_0t.
 \end{equation}
Equation (\ref{ele}) becomes:
\begin{equation}
B^2_*=B^2_{*0}\left(1+\frac{\beta-\alpha}{\beta+\alpha}\cos(2\chi_0+2\dot{\chi}_0t)\right).\label{twelve}
\end{equation}

Suppose the change of $\chi$ is small, thus Equation (\ref{twelve}) can be expanded as:
\begin{equation}
B^2_*=B^2_{*0}\left(1-2\gamma\dot{\chi}_0t\sin(2\chi_0)-2\gamma\dot{\chi}_0^2t^2\cos(2\chi_0)\right).\label{13}
\end{equation}
In Equation (\ref{13}), $B^2_{*0}$ is once again redefined to absorb the factor $1+(\beta-\alpha)/(\beta+\alpha)\cos(2\chi_0)$, and
\begin{equation}
\gamma\equiv\frac{(\beta-\alpha)/(\beta+\alpha)}{1+(\beta-\alpha)/(\beta+\alpha)\cos(2\chi_0)},
 \end{equation}
 which is a time-independent constant. Comparing with Equation (\ref{four}) we obtain:
 \begin{equation}
 b(t)=-\gamma\dot{\chi}_0t\sin(2\chi_0)-\gamma\dot{\chi}_0^2t^2\cos(2\chi_0).
 \end{equation}
 As a result, from Equation (\ref{five}) we expect to see the fourth-order polynomial as the pulse arrival phase:
 \begin{equation}
 \Phi(t)=\Phi_0+\nu_0t+\dot{\nu}_0\frac{t^2}{2}+\ddot{\nu}_0\frac{t^3}{3!}+\dddot{\nu}_0\frac{t^4}{4!}\label{poa},
\end{equation}
where
\begin{equation}
\begin{split}
&\ddot{\nu}_0=3\frac{\dot{\nu}_0^2}{\nu_0}-2\dot{\nu}_0\gamma\dot{\chi}_0\sin(2\chi_0),\\
&\dddot{\nu}_0=-4\dot{\nu}_0\gamma\dot{\chi}_0^2\cos(2\chi_0).
\end{split}\label{14}
\end{equation}

If we use the concept of characteristic age, $\tau\equiv-\nu_0/2\dot{\nu}_0$, then the first equation in Equations (\ref{14}) can be written as
\begin{equation}
\ddot{\nu}_0=-\dot{\nu}_0(\frac{3}{2\tau}+2\gamma\dot{\chi}_0\sin(2\chi_0)). \label{a}
\end{equation}

Several conclusions \textbf{can be drawn} from Equation (\ref{a}): a) for young pulsars whose $\tau$ are small, the first term in the bracket of Equation (\ref{a}) dominates and therefore have $\ddot{\nu}_0>0$; b) for old pulsars, the second term in the bracket of Equation (\ref{a}) dominate. Since $\dot{\chi}_0\sin(2\chi_0)$ can be either positive and negative, $\ddot{\nu}_0$ can be either $>0$ or $<0$; c): the absolute values of $\ddot{\nu}_0$ are proportional to $\dot{\nu}_0$ statistically, as long as the quantity $\gamma\dot{\chi}_0\sin(2\chi_0)$ distributes in a small range for all the pulsars. All of the three predictions have been confirmed by observations \citep[see][Figure1]{ZhangXie2012}.

The braking index is defined as $n=\ddot{\nu}\nu/{\dot{\nu}}^2$. From Equation (\ref{14}) we know that:
\begin{equation}
n-3=-\frac{2\nu_0}{\dot{\nu}_0}\gamma\dot{\chi}_0\sin(2\chi_0).\label{BI}
\end{equation}
Since $n$ is widely distributed over eight order of magnitude \citep[see][Figure 10]{ZhangXie2012}), we can infer that $\dot{\chi}_0\neq$ for most of the pulsars.

Equation (\ref{BI}) can also be rewritten in terms of the characteristic age:
\begin{equation}
n=3+4\tau\gamma\dot{\chi}_0\sin(2\chi_0).\label{BI2}
\end{equation}
Equation (\ref{BI2}) can be used to explain the observation fact that the absolute values of $n$ are proportional to $\tau$ \citep[see][Figure 2]{ZhangXie2012b}.

\begin{figure}
\centering
\includegraphics[width=9 cm]{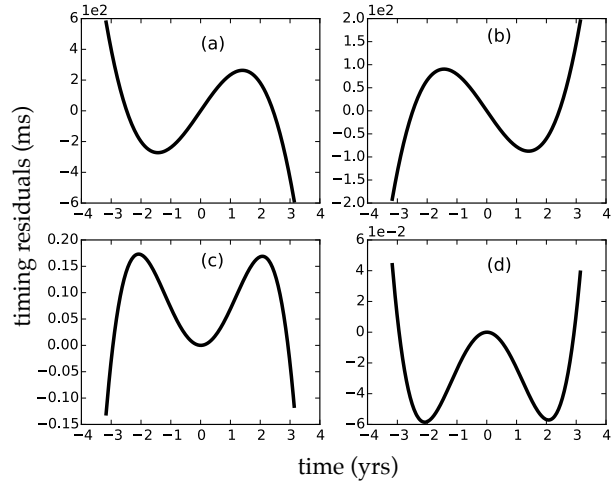}
\caption{Timing residuals of simulated TOAs after fitting a polynomial to the second degree. In all the simulation, $\nu_0=30.0$\,Hz, $\dot{\nu}_0=-3\times10^{-11}$\,s$^{-2}$ and $\gamma=1. $\textbf{(a)}: $\chi_0=0.75$, $\dot{\chi}_0=3\times10^{-12}$\,s$^{-1}$; \textbf{(b)}: $\chi_0=0.75$, $\dot{\chi}_0=-3\times10^{-12}$\,s$^{-1}$; \textbf{(c)}: $\chi_0=0.1$, $\dot{\chi}_0=-7.55\times10^{-12}$\,s$^{-1}$; \textbf{(d)}: $\chi_0=1.4$, $\dot{\chi}_0=-4.47\times10^{-12}$\,s$^{-1}$.}\label{figure1}
\end{figure}

We use Equation (\ref{poa}) and (\ref{14}) to simulate the TOAs, and fit a second degree polynomial to the TOAs. The resulting timing residuals are plotted in Figure \ref{figure1}, under different simulating parameters (see the caption of the figure). The shapes of four panels of Figure \ref{figure1} include most of that of the pulsars' timing residuals in the sample of \cite{Hobbs2010}. The variety of the shapes of observed timing residuals is due to various values of $\chi_0$ and $\dot{\chi}_0$ at different stages of the evolution.

\section{The origin of the $\chi$ evolution and the braking index}\label{section4}
The evolution of $\chi$ can come from either the intrinsic change of the magnetic field originating from the interior of the neutron star, or a kinematic process due to the rotation of the pulsar.
As mentioned above, both the magnetic dipole braking and the current loss predict a kinematic $\chi$ evolution, and the combination of them gives \citep{Barsukov2009}:
\begin{equation}
\dot{\chi}=\frac{AB_0}{2}\nu^2(\beta-\alpha)\sin(2\chi).\label{15}
\end{equation}
Therefore,
\begin{equation}
\dot{\chi}_0=\frac{AB_0}{2}\nu^2_0(\beta-\alpha)\sin(2\chi_0).\label{aaaa}
\end{equation}
From Equation (\ref{aaaa}) we see that if $\beta>\alpha$ then the magnetic dipole moves towards the equator of the pulsar, and when $\alpha>\beta$ the magnetic dipole evolves towards alignment with the spin axis.

In Equation (\ref{BI2}), if $\dot{\chi}_0$ is determined by Equation (\ref{aaaa}), then
\begin{equation}
\gamma\dot{\chi}_0\sin(2\chi_0)=\frac{AB_0}{2}\frac{(\beta-\alpha)^2/(\beta+\alpha)}{1+(\beta-\alpha)/(\beta+\alpha)\cos(2\chi_0)}\sin^22\chi_0.\label{xiongxiong}
\end{equation}

Therefore, as long as $\beta$ is positive (as widely supposed), $\gamma\dot{\chi}_0\sin(2\chi)_0>0$ and $n$ is aways larger than 3. It is intuitive since the kinematic processes will simultaneously adjust the configuration of the pulsar so that the braking torque decreases. As a result, a pulsar with $n<3$ requires that the intrinsic magnetic field of the neutron star is evolving. If we do not assume $\beta>0$, then $n<3$ requires that (from Equation (\ref{xiongxiong}))
\begin{equation}
\frac{\alpha+\beta}{\alpha-\beta}<\cos(2\chi_0).
\end{equation}
\section{Discussion}
\subsection{The magnetic inclination angle}
The magnetic inclination angle $\chi$ in this work is defined as the angle between the spin vector of the pulsar and the north pole of the magnetic dipole. This is different from the magnetic inclination angles $\chi_{\rm{ob}}$ which are measured by fitting the Rotation Vector Model (RVM) \citep{Radhak1969,Rookyard2015}. $\chi_{\rm{ob}}$ is defined as the angle between the spin vector and the end of magnetic dipole which is being observed. Therefore $\chi$ can be either equal to $\chi_{\rm{ob}}$ or $180^\circ-\chi_{\rm{ob}}$.

The same ambiguity also exists in $\dot{\chi}$: the observation can only tell whether the magnetic north pole is moving towards the spin axis or towards the equator, but can not tell whether $\chi$ is increasing or decreasing.

From Equation (\ref{14}) we have:
\begin{equation}
\frac{\dddot{\nu}_0}{\ddot{\nu}_0-3\dot{\nu}^2_0/\nu_0}=\xi\equiv2\dot{\chi}_0\cot(2\chi_0).\label{19}
\end{equation}
We denote $\xi$ which is calculated using the left-hand side of Equation (\ref{19}) as $\xi_{\rm{break}}$, and $\xi$ which is calculated by the definition in the right-hand side of Equation (\ref{19}) as $\xi_{\chi}$.
We use Equation (\ref{19}) and the parameters from the ATNF catalogue\footnote{http://www.atnf.csiro.au/research/pulsar/psrcat} \citep{Manchester2005} to calculate the values of $\xi_{\rm{break}}$, which are listed in Table {\ref{table}}. Other ATNF pulsars which are not listed here have no significant $\dddot{\nu}$. For a typical value of $\chi_0$=1\,rad, which is the expectation value assuming the isotropic distribution, $\cot(2\chi_0)\sim-0.46$ and $|\dot{\chi}_0|\sim|\xi|$. The estimated distribution of $\chi$ from observation depends heavily on the assumption on the shape of the radio beam. Therefore according to the values of $\dot{\chi}_0$ implied by Table \ref{table}, $\chi$ will approach to $0^\circ$ or $90^\circ$ in a time scale from years to thousands of years. These time scales are several orders of magnitudes less than the ages of pulsars, and based on the above section we expect that $\dot{\chi}_0\neq0$ for most of the pulsars. This implies that during the life time of a pulsar, $\dot{\chi}$  must change its sign for many times, which explains why there are approximately equal number of positive and negative $\xi$ in Table \ref{table}. The oscillation of $\chi$ necessarily causes oscillation of the equivalent $B$ of a pulsar, and thus might be responsible for the proposed oscillation of the magnetic field of neutron stars in \cite{ZhangXie2012,ZhangXie2012b}.
\begin{table}
\centering
\begin{tabular}{ll}
\hline
name & $\xi$\,(rad s$^{-1}$)\\
\hline
\hline
J0007+7303 & $1.89\times10^{-7}$\\
J0534+2200 & $2.70\times10^{-10}$\\
J1023-5746 & $-1.61\times10^{-8}$\\
J1418-6058 & $-1.39\times10^{-7}$\\
J1513-5908 & $8.0\times10^{-10}$\\
J1623-2631 & $3.31\times10^{-10}$\\
J1824-2452A & $-9.92\times10^{-8}$\\
J1833-0831 & $-6.92\times10^{-8}$\\
J2337+6151 & $-2.73\times10^{-8}$\\
\hline
\end{tabular}
\caption{\textbf{The values of $\xi_{\rm{break}}$ of ANTF pulsars.}}\label{table}
\end{table}

\subsection{The timing residuals of the Crab pulsar}
\cite{Lyne2013} used the inferred drift of $\chi$ to explain the braking index of 2.5 of the Crab pulsar (PSR J0534+2200). From Table \ref{table} we see that $\xi_{\rm{break}}=2.70\times10^{-10}$\,rad s$^{-1}$ for the Crab pulsar. However the observation suggests that $\dot{\chi}_0=\pm3\times10^{-12}$ rad s$^{-1}$ towards the equator, and $45^\circ<\chi_0<70^\circ$ or $110^\circ<\chi_0<135^\circ$\citep{Dyks2003,Harding2008,Watters2009,Du2012}. As a result, we have $-7\times10^{-12}<\xi_{\chi}<0$\,rad s$^{-1}$.  The inconsistency between $\xi_{\rm{break}}$ and $\xi_{\chi}$ indicates that more evolution of the neutron star magnetic field is required for the rotation behavior of the Crab pulsar. From section \ref{section4} we learnt that since the braking index of the Crab pulsar is less than 3, an intrinsic change of the magnetic field is needed. Therefore the strength and $\chi$ of the magnetic field of the Crab pulsar might evolve together. Thus, the $B$ field strength evolution mechanisms \citep{Lin2004,Chen2006,Espinoza2013,ZhangXie2012,ZhangXie2012b} can not be replaced by the pure $\chi$ evolution. The pulsar wind braking \citep{Xu2001,Wu2003,Contopoulos2006} is another competitive explanation for the abnormal rotational behavior of the Crab pulsar, which has been studied by \cite{Kou2015}.

\subsection{Correction for the red timing-noise}
The evolution of $\chi$ induces the extra cubic and fourth-order terms in the timing residuals, therefore a fitting to the fourth-order polynomial can remove the effects. However an arbitrary fitting of polynomial may also remove other useful signals in the timing residuals, e.g., GW. Although $\chi$ and $\dot{\chi}_0$ can be constrained by observations, we can not yet constrain $\ddot{\nu}_0$ and $\dddot{\nu}_0$ since $\gamma$ is unknown in Equations (\ref{14}). Nevertheless, these two terms are coupled with each other by Equation (\ref{19}).
Therefore if $\chi$ and $\dot{\chi}$ (and thus $\xi_{\chi}$) are determined by independent methods, we can reduce the number of free parameters as shown in
\begin{equation}
 \Phi(t)=\Phi_0+\nu_0t+\dot{\nu}_0\frac{t^2}{2}+\ddot{\nu}_0\frac{t^3}{3!}+\xi_{\chi}(\ddot{\nu}_0-3\frac{\dot{\nu}_0^2}{\nu_0})\frac{t^4}{4!}\label{poa2},
\end{equation}
and thus alleviate the removal of other useful signals while fitting a fourth-order polynomial to the timing residuals.
\section{Summary and Conclusion}
We modeled the braking torque of the pulsar spin-down as a combination of the dipole radiation and the current loss, and derived the red timing-noise caused by the evolution of the magnetic inclination angle $\chi$ of pulsars. With this model, we reproduced the four typical types of timing residuals observed in the sample of \cite{Hobbs2010}. Comparing with observations, we calculated the quantity $2\dot{\chi}\cot(2\chi)$ for nine pulsars in the ATNF catalogue.

Our conclusions are as follows:
\begin{enumerate}
\item{The cubic and the fourth-order polynomial terms in the timing residuals can be explained as a result of the evolution of magnetic inclination angle (see Equations (\ref{14})), if we consider the spin-down mechanism of pulsars to be a combination of the magnetic dipole radiation and the current loss. The variety of shapes of the timing residuals originate from the various values of $\chi$ and $\dot{\chi}$ at different stages of the evolution (see Figure \ref{figure1}).}
\item{The evolution of magnetic inclination angle can explain the following observation facts: a) young pulsars (small $\tau$) have $\ddot{\nu}_0>0$; b) old pulsars can have either $\ddot{\nu}_0<0$ or $\ddot{\nu}_0>0$; c) $|\ddot{\nu}_0|$ are proportional to $-\dot{\nu}_0$ among pulsars; d) $|n|$ are proportional to $\tau$.
}
\item{The evolution of $\chi$ purely due to rotation kinematics can not explain the pulsars with braking index less than 3. The intrinsic change of the magnetic field (either strength or $\chi$) is needed.}
\item{The sign of $\dot{\chi}$ changes many times during the lifetime of a pulsar, which might be responsible for the proposed oscillation of the magnetic field of neutron stars in \cite{ZhangXie2012,ZhangXie2012b}.}
\item{The evolution of $\chi$ is not sufficient to explain the rotation behavior of the Crab pulsar.}
\end{enumerate}

\section*{acknowledgement}
SXY thanks the helpful discussion with Dr. Rookyard about the observation of the magnetic field inclination angle. SNZ acknowledges
partial funding support by 973 Programof China under grant 2009CB824800, by theNational Natural Science Foundation of China under grant Nos. 11133002, 11373036 and 10725313, and by the Qianren start-up grant 292012312D1117210

\bibliographystyle{mn2e}

\begin{thebibliography}{1}
\bibitem[Barsukov et al.(2009)]{Barsukov2009} Barsukov, D.~P.,
Polyakova, P.~I., \& Tsygan, A.~I.\ 2009, Astronomy Reports, 53, 1146
\bibitem[Beskin et al.(1993)]{Beskin1993} Beskin, V.~S., Gurevich,
A.~V., \& Istomin, Y.~N.\,Physics of the pulsar magnetosphere, Cambridge University Press (1993)
\bibitem[Beskin
\& Nokhrina(2007)]{Beskin2007} Beskin, V.~S., \& Nokhrina, E.~E.\ 2007, Astrophysics and Space Science, 308, 569
\bibitem[Boynton et al.(1972)]{Boynton1972} Boynton, P.~E., Groth,
E.~J., Hutchinson, D.~P., et al.\ 1972, ApJ, 175, 217
\bibitem[Contopoulos
\& Spitkovsky(2006)]{Contopoulos2006} Contopoulos, I., \& Spitkovsky, A.\ 2006, ApJ, 643, 1139
\bibitem[Chen
\& Li(2006)]{Chen2006} Chen, W.~C., \& Li, X.~D.\ 2006, A\&A, 450, L1
\bibitem[Detweiler(1979)]{Detweiler1979} Detweiler, S.\ 1979, ApJ,
234, 1100
\bibitem[Du et al.(2012)]{Du2012} Du, Y.~J., Qiao, G.~J.,
\& Wang, W.\ 2012, ApJ, 748, 84
\bibitem[Dyks
\& Rudak(2003)]{Dyks2003} Dyks, J., \& Rudak, B.\ 2003, ApJ, 598, 1201
\bibitem[Espinoza(2013)]{Espinoza2013} Espinoza, C.~M.\ 2013, IAU
Symposium, 291, 195
\bibitem[Haensel et
al.(1990)]{Haensel1990} Haensel, P., Urpin, V.~A., \& Iakovlev, D.~G.\ 1990, A\&A, 229, 133
\bibitem[Harding et al.(2008)]{Harding2008} Harding, A.~K., Stern,
J.~V., Dyks, J., \& Frackowiak, M.\ 2008, ApJ, 680, 1378
\bibitem[Hobbs et al.(2010)]{Hobbs2010} Hobbs, G., Lyne, A.~G.,
\& Kramer, M.\ 2010, MNRAS, 402, 1027
\bibitem[Jones(1976)]{Jones1976} Jones, P.~B.\ 1976, ApJ, 209,
602
\bibitem[Jones(1990)]{Jones1990} Jones, P.~B.\ 1990, MNRAS, 246,
364
\bibitem[Kou
\& Tong(2015)]{Kou2015} Kou, F.~F., \& Tong, H.\ 2015, MNRAS, 450, 1990
\bibitem[Kramer et al.(2006)]{Kramer2006} Kramer, M., Lyne, A.~G.,
O'Brien, J.~T., Jordan, C.~A., \& Lorimer, D.~R.\ 2006, Science, 312, 549
\bibitem[Lin
\& Zhang(2004)]{Lin2004} Lin, J.~R., \& Zhang, S.~N.\ 2004, ApJL, 615, L133
\bibitem[Lyne et al.(2010)]{Lyne2010} Lyne, A., Hobbs, G.,
Kramer, M., Stairs, I., \& Stappers, B.\ 2010, Science, 329, 408
\bibitem[Lyne et al.(2013)]{Lyne2013} Lyne, A., Graham-Smith,
F., Weltevrede, P., et al.\ 2013, Science, 342, 598
\bibitem[Lyne et al.(2015)]{Lyne2015} Lyne, A.~G., Jordan,
C.~A., Graham-Smith, F., et al.\ 2015, MNRAS, 446, 857
\bibitem[Manchester et al.(2005)]{Manchester2005} Manchester, R.~N.,
Hobbs, G.~B., Teoh, A., \& Hobbs, M.\ 2005, The Astronomical Journal, 129, 1993
\bibitem[Mestel et al.(1999)]{Mestel1999} Mestel, L., Panagi, P.,
\& Shibata, S.\ 1999, MNRAS, 309, 388
\bibitem[Radhakrishnan
\& Cooke(1969)]{Radhak1969} Radhakrishnan, V., \& Cooke, D.~J.\ 1969, Astrophys. Lett., 3, 225
\bibitem[Rookyard et al.(2015)]{Rookyard2015} Rookyard, S.~C.,
Weltevrede, P., \& Johnston, S.\ 2015, MNRAS, 446, 3367
\bibitem[Sazhin(1978)]{Sazhin1978} Sazhin, M.~V.\ 1978, Soviet Astronomy,
22, 36
\bibitem[Shannon et al.(2013)]{Shannon2013} Shannon, R.~M., Cordes,
J.~M., Metcalfe, T.~S., et al.\ 2013, ApJ, 766, 5
\bibitem[Watters et al.(2009)]{Watters2009} Watters, K.~P., Romani,
R.~W., Weltevrede, P., \& Johnston, S.\ 2009, ApJ, 695, 1289
\bibitem[Wu et
al.(2003)]{Wu2003} Wu, F., Xu, R.~X., \& Gil, J.\ 2003, A\&A, 409, 641
\bibitem[Xu
\& Qiao(2001)]{Xu2001} Xu, R.~X., \& Qiao, G.~J.\ 2001, ApJL, 561, L85
\bibitem[Zhang
\& Xie(2012a)]{ZhangXie2012} Zhang, S.-N., \& Xie, Y.\ 2012a, ApJ, 757, 153
\bibitem[Zhang
\& Xie(2012b)]{ZhangXie2012b} Zhang, S.-N., \& Xie, Y.\ 2012b, ApJ, 761, 102

\end{thebibliography}

\end{document}